\documentclass[aps,prb,showpacs,showkeys,12pt,reprint]{revtex4-1}
\usepackage[letterpaper, margin=0.75in]{geometry}
\usepackage{amsmath} 
\usepackage[dvips]{graphicx}
\usepackage{tabularx}
\usepackage{float}
\usepackage{multirow}
\usepackage{ctable}
\usepackage{array}
\usepackage{notoccite}
\usepackage{setspace}
\usepackage{mathtools}
\usepackage[strict]{chngpage}

\graphicspath{{./}}
\DeclareGraphicsExtensions{.jpg}
\usepackage{rotating}

\newcommand{\thline}{\specialrule{.13em}{.05em}{.05em}}
\newcommand{\angbra}[1]{$\langle$#1$\rangle$}
\newcolumntype{Y}{>{\centering\arraybackslash}X}
\begin{document}
\title[\emph{Ab initio} based empirical potential applied to tungsten at high pressure]{\emph{Ab initio} based empirical potential applied to tungsten at high pressure}
\author{\rm Robert C. Ehemann$^{1}$}
\author{\rm Jeremy W. Nicklas$^{2}$}
\author{\rm Hyoungki Park$^{1}$}
\author{\rm John W. Wilkins$^{1}$}
\address{$^{1}$ Department of Physics, The Ohio State University, Columbus, Ohio 43210, USA}
\address{$^{2}$ The Ohio Supercomputer Center, Columbus, Ohio 43212, USA}
\begin{abstract}
	Density-functional theory forces, stresses and energies comprise a database from which the optimal parameters of a spline-based empirical potential combining Stillinger-Weber and modified embedded-atom forms are determined. Accuracy of the potential is demonstrated by predictions of ideal shear, stacking fault, vacancy migration, elastic constants and phonons all between 0 and 100 GPa. Consistency with existing models and experiments is demonstrated by application to screw dislocation core structure and deformation twinning in a tungsten nanorod. Lastly, the potential is used to study the high-pressure bcc to fcc phase transition.
\end{abstract}

\date{\today}
\maketitle

\section{Introduction}
Tungsten is an exceptional transition metal exhibiting the highest tensile strength, melting point, and elastic modulus of any pure metal and has important applications in aerospace, energy and armament industries. Much interest has been focused on $\alpha$-W (bcc) and $\beta$-W (A15) nanostructures including nanorods\cite{Karabacak2003, Karabacak2005, Wang2015}, nanoparticles\cite{Ozawa2001, Lei2007, Shibuta2008, Harilal2013}, and thin films\cite{Chopra1967, Haghri-Gosnet1989, Weerasekera1994}. Due to the technological importance of tungsten, classical interatomic potentials of various forms have been developed to study this metal\cite{Ackland1987,Baskes1992,Zhou2004,Derlet2007,Mrovec2007,Wang2014,Murphy2015}.

Classical potentials -- popular for their favorable scaling compared to first-principles methods ($\mathcal{O}(N)$ vs. $\sim\mathcal{O}(N^3)$) -- were traditionally developed by choosing analytic functional forms with a handful of free parameters determined by fitting directly to experimental bulk data such as cohesive energy, lattice and elastic constants. The force-matching method of Ercolessi and Adams\cite{Ercolessi1994} has facilitated the development of interatomic potentials based on \emph{ab initio} calculations of relaxed crystallographic defects, metastable structures and other non-equilibrium configurations. However, many such potentials still possess a small number of free parameters and analytic functional forms which limit their transferability, requiring researchers to take great caution in choosing a potential suitable for their region of interest. It is thus desirable to produce a single potential which can be employed to study the range of materials physics, here in tungsten.

To meet this challenge we develop an unique semi-empirical potential based on a robust database of \emph{ab initio} calculations that samples much of the potential-energy landscape. Our model combines the Stillinger-Weber (SW)\cite{Stillinger1985}  form with the modified embedded atom method\cite{Daw1984} (MEAM) form with functions parameterized by quintic splines. Section \ref{sec:opt} describes the functional form of the model, the density-functional theory (DFT) calculations comprising the large fitting database, and the genetic algorithm optimization scheme. Accuracy of the fitted potential is demonstrated in Section \ref{sec:acc} by comparing MEAM to DFT for the various structural and elastic properties to which it was fit. Given that the potential is fit directly to important crystallographic defects, structural properties and elastic constants, transferability is demonstrated in Section \ref{sec:tra} by examining MEAM predictions for $\frac{1}{2}$\angbra{111} screw dislocation core structure, deformation twinning and detwinning of a nanorod, and dynamics of bcc and fcc tungsten at high pressure. Conclusions are given in Section \ref{sec:con}.
\section{Optimization of the empirical potential with first-principles calculations} \label{sec:opt}
We present a spline-based empirical potential fit to a large database of highly-converged density functional theory (DFT) calculations using a global optimization scheme based on an evolutionary algorithm.
\subsection{Empirical extension of the MEAM potential}
The embedded-atom (EAM)\cite{Daw1984, Daw1993} and MEAM\cite{Baskes1987, Baskes1989, Baskes1992} methods have been applied to many systems including semiconductors\cite{Daw1984,Baskes1987, Baskes1989, Daw1993, Lenosky2000, Lee2007} and transition metals\cite{Baskes1992, Mishin1999, Hennig2008, Fellinger2010, Park2012}. The original MEAM formalism involves a parametrized analytical functional form which accounts for bond-bending through angular functions with explicit $s$-, $p$-, $d$- and $f$-orbital characteristics. Lenosky {\emph{et al.}}\cite{Lenosky2000} first parametrized the MEAM formalism through the use of cubic splines for the study of defects in Si. The use of splines for parameterizing empirical potentials increases model flexibility and efficiency, and has been successfully applied to the study of martensitic transformations in pure titanium\cite{Hennig2008}, shock-loading in niobium\cite{Fellinger2010, Zhang2011} and dislocation dynamics in molybdenum\cite{Park2012}. SW potentials, initially developed for the modeling of cubic-diamond Si, have been successfully applied to amorphous Si\cite{Vink2001} as well as Ge\cite{Jian1990} and other systems.

Model flexibility is paramount when constructing empirical potentials. The addition of distinct terms to the model can  improve flexibility, as demonstrated by the success of MEAM over EAM \cite{Cherne2001,Vella2015}. In the present work, we propose an empirical extension of MEAM which includes a SW-type three-body term in the energy. Our model employs functions parameterized by quintic splines which improve performance for properties requiring continuous third derivatives of the interatomic potential, e.g. $C_{ij}$ vs. $P$ relations, over cubic splines. The total potential energy is that of SW with the addition of an embedding term $U$,
\begin{equation} \label{eq:EMEAM}
\begin{split}
&V_{tot} = \sum\limits_{i>j}\phi(r_{ij}) + \sum\limits_{i}U(n_{i}) \\ &+ \sum\limits_{\substack{i,j>k\\j\neq i}}p(r_{ij})p(r_{ik})q(\mathrm{cos}\ \theta_{jik}),
\end{split}
\end{equation}
\noindent where the ``electronic density" $n_{i}$ at atom $i$ involves two- and three-body contributions
\begin{equation} \label{eq:nMEAM}
\begin{split}
n_{i} = \sum\limits_{j{\neq}i}\rho(r_{ij}) 
+ \sum\limits_{\substack{j>k\\j\neq i}}f(r_{ij})f(r_{ik})g(\mathrm{cos}\ \theta_{jik}),
\end{split}
\end{equation}
\noindent and $\theta_{jik}$ is the angle of the triplet centered on atom $i$. This is the simplest extension of MEAM which does not include four-body terms and cannot be gauge-transformed back to the original model. It contains as special cases the SW ($U=0$),  MEAM ($q=0$), and EAM ($g=q=0$) forms. This flexibility gives us the ability to fit to a large ab-initio database described in the next section.
\subsection{Density-functional theory database}
DFT calculations performed with {\sc{vasp}}\cite{Kresse1993, Kresse1994, Kresse1996a, Kresse1996b} using a projector-augmented planewave basis\cite{Blochl1994} and Perdew-Burke-Erzenhof (PAW-PBE)\cite{Perdew1996, Perdew1997} generalized-gradient exchange correlation approximation comprise a database of forces, stresses and energies for fitting via the force-matching method. In addition to 6$s$ and 5$d$, the 5$p$ electrons are treated as valence to improve accuracy at high pressures, where the overlap of these semi-core states is not necessarily negligible. A planewave energy cutoff of 600 eV and first-order Methfessel-Paxton smearing width of 0.1 eV are used for all calculations; k-points are sampled on a $\Gamma$-centered 40$\times$40$\times$40 mesh in the bcc brillouin zone. These quantities are chosen to ensure convergence of the total energy to within 0.1 meV/atom. Additional computational details are presented in Fellinger\cite{Fellinger2013}.

The \emph{ab-initio} fitting database contains 596 configurations with a total of 16,860 atoms and thus 54,752 force components, stress components and energies to be fit. Configurations in the database include volumetric strains of bcc, fcc, hcp, $\beta$-W (A15), $\beta$-Ta ($\beta$-U) and $\omega$-Ti. Tetragonal strains are included for hcp and $\omega$-Ti structures to ensure accurate $c/a$ values. The database also contains elastic constants of the bcc phase at pressures between 0 and 100 GPa, in increments of 25 GPa, using volume-conserving orthorhombic and monoclinic strains of 0.5 \% for $C^{\prime} = \frac{1}{2}(C_{11}-C_{12})$ and $C_{44}$, respectively. At zero pressure, configurations with orthorhombic strains up to 10 \% and monoclinic strains up to 40 \% are added. Unrelaxed symmetry-inequivalent configurations of \angbra{110} and \angbra{112} $\gamma$-surfaces, ideal shear strain and vacancy migration are included at five equally spaced pressures between 0 and 100 GPa. Relaxed zero-pressure structures containing a vacancy at the lattice site and halfway along the \angbra{111} migration path are also added. A 7$\times$7$\times$7 bcc supercell with a single atom displaced by 0.006 $\mathrm{\AA}$ is included to promote accurate force-constants and phonons via the small-displacement method\cite{Kresse1995, Alfe2001}. Using a supercell of this size reduces the interaction of the displaced atom with its images across periodic boundaries and thus improves the accuracy of calculated force-constants and phonon dispersions. Relaxed low-index free surfaces as well as crowdion, octahedral, \angbra{111}-split and \angbra{110}-split self-interstitial configurations are included. \emph{Ab-initio} MD snapshots of 125-atom bcc supercells at 1620 K, 2960 K and a liquid tungsten at 6730 K are added to improve performance for simulations at high temperature. A 36-atom hcp supercell at 100 K is also included. A Lastly, a mesh of 36 points on the Bain\cite{Bain1924} (bcc-fcc) and Burgers\cite{Burgers1934} (bcc-hcp) energy surfaces at pressures of 0 GPa and 700 GPa in addition to 600 GPa for the Bain path and 800 GPa for the Burgers path are included to ensure that the potential can be used to explore properties of these close-packed phases at high pressure.
\subsection{Genetic algorithm optimization}
Development of the optimized potential is an iterative process of fitting, testing and database refinement. Ten to twenty fits are performed simultaneously and the resultant potentials are tested for accuracy on a range of properties. The fitting database is refined based on the results of these tests: new structures are added to correct spurious behavior and structures are removed when over-fitting is suspected. Re-tuning of algorithm inputs and/or error weights often accompanies this refinement.

A global optimization scheme combining a genetic algorithm (GA) with a local downhill optimizer provides a method for determining the spline parameters of the potential. At each iteration of the GA all potentials in the population of ten are locally optimized with 60 steps of a Powell\cite{Powell1964} conjugate direction algorithm, then the population is sorted and bred according to total weighted least-squares error. For the presented potential forces, stresses and energies are equally weighted. Breeding is done by a stochastic combination of spline knots from two parent potentials. The following constraints are enforced by introducing a ``punishment'' error when not satisfied: (i) $\left|\mathrm{max}[f(r)]\right| = \left|\mathrm{max}[p(r)]\right| = 1$ and (ii) $n_i$ lies within the domain of $U(n)$ for all $i$. If the latter constraint is violated, the embedding function is evaluated at its nearest endpoint.  At each GA step, for every potential in the population, there is a 10\% chance for the embedding function domain and total density (Eq. \ref{eq:nMEAM}) to be rescaled by a transformation where $U(n) \rightarrow U(n/\alpha)$ and $n \rightarrow \alpha n$, where $\alpha$ is determined by the minimum and maximum densities at the current step. When this occurs, additional gauge symmetries in the three-body terms of Eqs. \ref{eq:EMEAM} and \ref{eq:nMEAM} are exploited so that the maximal knot values of $\left|f\right|$ and $\left|p\right|$ are equal to 1.

While forces and energies are invariant with respect to these transformations, the total error is not because constraints (i) and (ii) are always satisfied after rescaling. Because during fitting the embedding function is not extrapolated but rather evaluated at the nearest endpoint when densities lie outside the domain, energies for such configurations are not invariant under the aforementioned rescaling. Furthermore, spline functions are in general not invariant under such a transformation of their argument. Thus, performing this rescaling serves as a genetic mutation of the potential. The algorithm is exited when between successive steps the change in total error for every potential in the population is less than $10^{-3}$. Parameters for the fitted potential and plots of the seven functions are presented in the Supplementary Information, along with more detailed descriptions of the algorithms used. Henceforth all references to MEAM will pertain to the present empirically extended potential.
\section{Accuracy of the fitted potential} \label{sec:acc}
We demonstrate the accuracy of the fitted MEAM potential through the energetics of non-equilibrium structures, crystallographic defects, thermodynamic properties and phonon dispersion. All MEAM calculations in this work (other than those necessary for fitting) are performed in the Large-scale Atomic/Molecular Massively Parallel Simulator ({\sc{lammps}})\cite{Plimpton1995}. If at any step during an MD run the density seen by an atom exceeds the embedding function domain, the embedding energy is linearly extrapolated from the nearest endpoint.
\subsection{Energetics and elastic properties}
\begin{figure}[H]
	\begin{center}
		\includegraphics[width=0.47\textwidth]{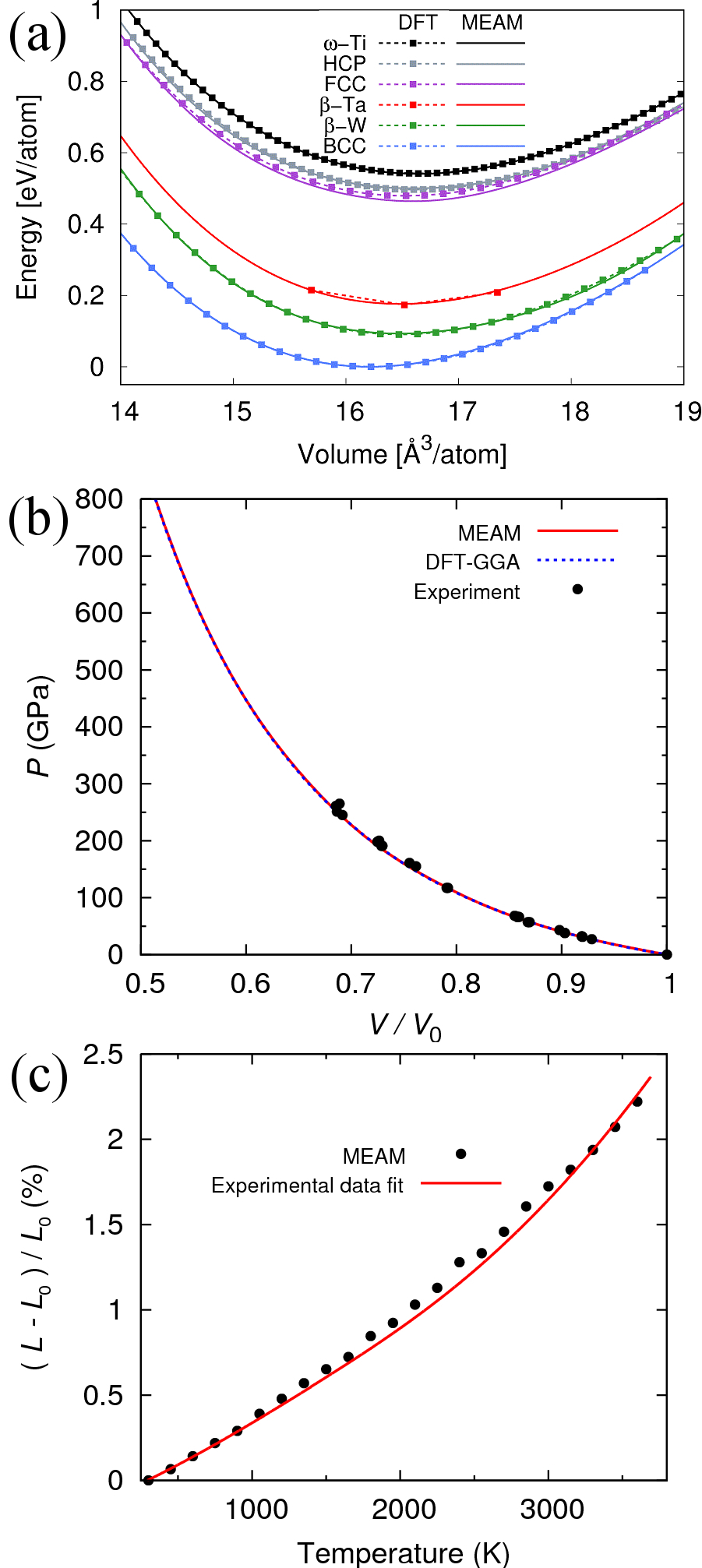}
		\caption{(a) Comparison of energy-volume curves between MEAM and DFT for six crystal structures. Curves are ordered vertically according to the key. Our empirical potential reproduces the energies of each phase relative to that of the bcc ground state. (b) Pressure-volume relation for tungsten as computed by MEAM and DFT compared with data from shock experiments\cite{Kinslow1970}. MEAM shows agreement with both experimental and \emph{ab-initio} results, even at extreme pressures. (c) Thermal expansion of tungsten predicted by MEAM agrees with experimental data fit\cite{Touloukian1975} up to the melting point of 3695 K.} \label{fig:evpvtv}
	\end{center}
\end{figure}

Figure \ref{fig:evpvtv}(a) shows the GGA-DFT and MEAM energy-volume relations for six distinct phases including A15 $\beta$-W and high-energy close-packed structures. MEAM accurately predicts energies of all six phases relative to the ground state. An equilibrium bcc lattice constant of 3.189 $\mathrm{\AA}$ is predicted by both GGA-DFT and MEAM, compared to the published experimental values between 3.15 and 3.165 $\mathrm{\AA}$\cite{Davey1925,Hartmann1931,Taylor1967}. It is well known now that GGA tends to overestimate the lattice constant of metals; the reason for this is discussed in Wang \emph{et al.}\cite{Wang2004} as well as Favot and Dal Corso\cite{Favot1999} and references therein.
Figure \ref{fig:evpvtv}(b) compares bcc tungsten pressure-volume relations as computed with MEAM, GGA-DFT, and measured through shock experiments\cite{Kinslow1970}. MEAM and GGA-DFT curves, obtained by static calculations of volumetric strain, are indistinguishable for pressures through 800 GPa and in excellent agreement with experimental results up to 300 GPa, indicating applicability of the fitted MEAM potential to high-pressure physics in tungsten.

\begin{figure}[H]
	\begin{center}
		\includegraphics[width=0.47\textwidth]{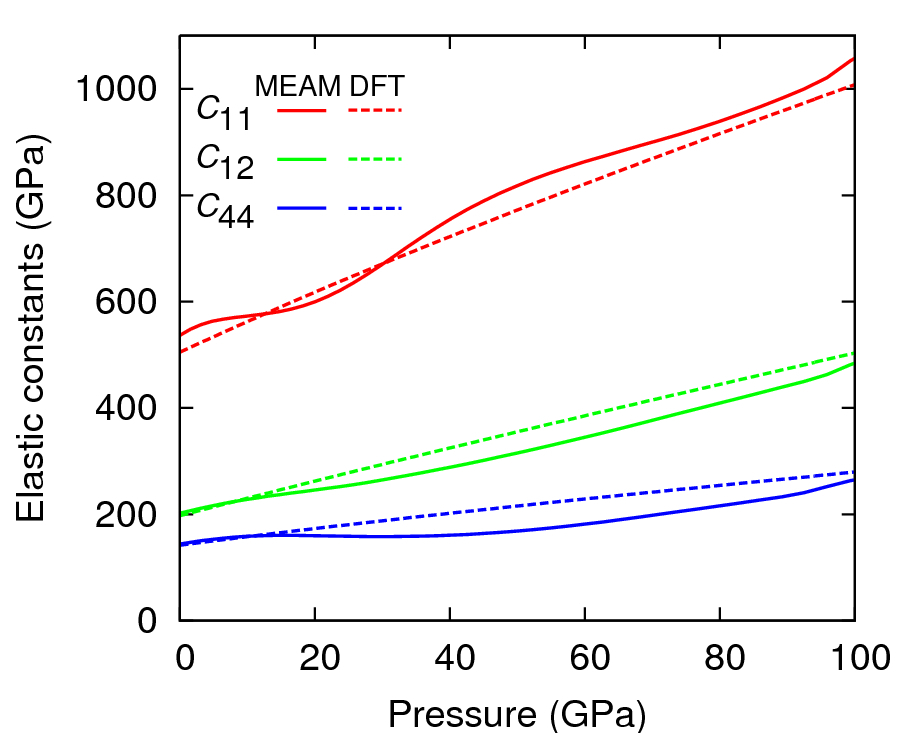} 
		\caption{Elastic constants versus pressure for bcc tungsten as computed with GGA-DFT and the fitted potential. MEAM produces elastic constants within 21\% of the DFT values for all pressures shown.} \label{fig:CvsP}
	\end{center}
\end{figure}

\begin{figure}[H]
	\begin{center}
		\includegraphics[width=0.5\textwidth]{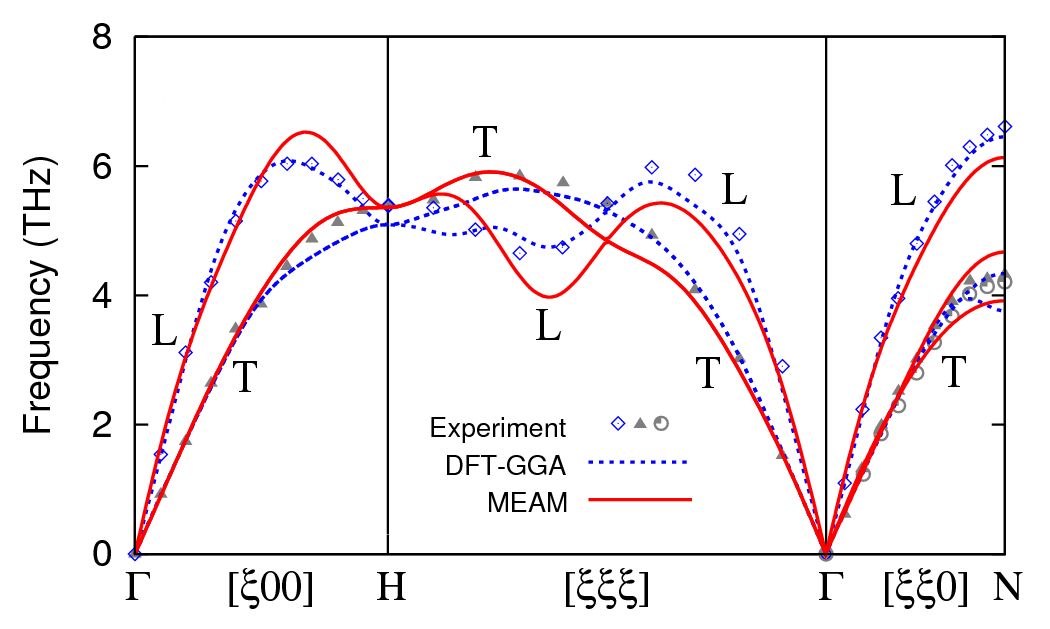}
		\caption{Phonon dispersion for bcc W at zero pressure as calculated by DFT and MEAM, compared with inelastic neutron scattering data of Chen and Brockhouse\cite{Chen1964}.} \label{fig:Phonons}
	\end{center}
\end{figure}

Figure \ref{fig:evpvtv}(c) compares linear thermal expansion predictions by MEAM to experimental results\cite{Touloukian1975} for temperatures between 300 K and the experimental melting point of 3695 K. Constant N-P-T MD simulations of 2000 atoms at P = 1 atm yield the thermal-expansion curve. Each MD simulation runs for 50,000 steps with a 1 fs timestep and the lattice constant for each temperature is determined by averaging over the last 5,000 simulation steps. MEAM shows excellent agreement with experiment up to 1,000 K and remains within 1 \% of the experimental fit for all temperatures considered, indicating that the potential interpolates between temperatures included in the fitting database. 
\begin{table}[t]
	\caption{Zero-pressure elastic constants of bcc tungsten in GPa.} \label{tab:elcons}
	\begin{tabularx}{0.5\textwidth}{l Y Y Y Y}
		\hline
		\thline       & B        & C$_{11}$ & C$_{12}$ & C$_{44}$ \\ \hline
		MEAM$^{a}$  & 319      & 550      & 204      & 147      \\
		GGA$^{a}$     & 304      & 513      & 199      & 142      \\
		F-S$^{b}$     & 309      & 520      & 204      & 161      \\
		LDA$^{c}$     & 320      & 552      & 204      & 149      \\
		F-S$^{d}$     & 310      & 525      & 203      & 159      \\
		F-S$^{e}$     & 310      & 522      & 204      & 161      \\
		EAM$^{f}$     & 308      & 520      & 202      & 159      \\
		BOP$^{g}$     & 310      & 522      & 204      & 161      \\
		Expt.$^{h}$   & 308--314 & 501--521 & 199-207  & 151--160 \\
		\thline\hline 
	\end{tabularx}
	\begin{tabularx}{0.5\textwidth}{l}
	$^{a}$MEAM and GGA-DFT results of this work. \\
	$^{b}$Finnis-Sinclair results of Wang \emph{et al.}\cite{Wang2014} \\
	$^{c}$LDA-DFT results of Einarsdotter \emph{et al.}\cite{Einarsdotter1997}\\
	$^{d}$Finnis-Sinclair results of Derlet \emph{et al.}\cite{Derlet2007} \\
	$^{e}$Finnis-Sinclair results of Ackland \emph{et al.}\cite{Ackland1987} \\
	$^{f}$EAM results of Zhou \emph{et al.}\cite{Zhou2004} \\
	$^{g}$Bond-order potential results of Mrovec \emph{et al.}\cite{Mrovec2007} \\
	$^{h}$Expt. results of Bolef \emph{et al.} from 77 to 500 K\cite{Bolef1962}
	\end{tabularx}
\end{table}
Table \ref{tab:elcons} shows the zero-pressure bcc elastic constants of the present MEAM and GGA-DFT results, compared to previous \emph{ab initio} calculations and other interatomic potentials. The bulk modulus B and C$_{11}$ predicted by MEAM are higher than experimental and GGA results but consistent with the LDA work of Einarsdotter \emph{et al.}\cite{Einarsdotter1997}. The pressure-dependence of bcc elastic constants is shown in Figure \ref{fig:CvsP}; MEAM does not predict a monotonic increase of C$_{ij}$ but remains within 21 \% of the GGA-DFT values.
Figure \ref{fig:Phonons} shows phonon dispersion of equilibrium bcc tungsten as computed with MEAM and DFT, compared to inelastic neutron scattering results of Chen and Brockhouse\cite{Chen1964}. Dispersions are calculated using the finite-displacement method in a 7$\times$7$\times$7 primitive bcc supercell. DFT dispersion agrees well with experiment but exhibits oscillations in the longitudinal (low-lying) branch near the $H$-point, a feature also found in density-functional perturbation theory results within LDA-DFT\cite{Einarsdotter1997}. Overall MEAM tracks both DFT and experiment but underestimates the frequency along the L[$\xi\xi\xi$] branch, particularly near the $\omega$ mode at $\xi = \frac{2}{3}$.
\begin{table*}[t!]
	\caption{Table of vacancy and self-interstitial formation energies (in eV) for bcc tungsten. Entries with angle-brackets indicate that the defect in question relaxes to the dumbell configuration shown.} \label{tab:purepds}
	\begin{tabularx}{\textwidth}{l Y Y Y Y Y Y Y Y}
		\hline
		\thline
		Defect     & MEAM$^{a}$ & GGA$^{a}$ & F-S$^{b}$ & GGA$^{c}$ & F-S$^{d}$ & F-S$^{e}$ & EAM$^{f}$ & GGA$^{g}$   \\ \hline
		Vac. Formation    & 2.99          & 3.17          & 3.58      & 3.11           & 3.56      & 3.63      & 3.57  & 3.56      \\
		Vac. Migration    & 1.73$^{\dagger}$          & 1.70$^{\dagger}$          & 1.43      & 1.66           & 2.07      & 1.44      & 2.98$^{\dagger}$   & 1.78       \\
		Vac. Activation   & 4.72$^{\dagger}$          & 4.87$^{\dagger}$          & 5.01      & 4.77           & 5.63      & 5.07      & 6.55$^{\dagger}$  & 5.34       \\
		\angbra{001} Dumbell & 11.15         & \angbra{111}  & 11.53     & 11.74           & 11.51     & 9.82      & 12.20 & 11.49       \\
		\angbra{011} Dumbell & 9.98          & 10.64         & 9.86      & 10.10           & 9.84      & 9.64      & 9.704 & 9.84      \\
		\angbra{111} Dumbell & 9.73          & 10.31         & 9.58      & 9.82           & 9.55      & 9.82      & 10.56  & 9.55     \\
		Octahedral           & 11.76         & 12.42         & 11.72     & 11.99           & 11.71     & 10.02     & 12.03 &  11.68     \\
		Tetrahedral          & 10.54         & \angbra{111}  & 10.93     & 11.64           & 11.00     & 10.00     & \angbra{011} & 11.05\\
		\thline\hline        
	\end{tabularx}
	\begin{tabularx}{\textwidth}{l | l l}
		$^{\dagger}$Unrelaxed calculation by present authors. & &
		$^{d}$Finnis-Sinclair results of Derlet \emph{et al.}\cite{Derlet2007} \\
		$^{a}$MEAM and GGA-DFT results of this work. & &
		$^{e}$Finnis-Sinclair results of Ackland \emph{et al.}\cite{Ackland1987} \\	
		$^{b}$Finnis-Sinclair results of Wang \emph{et al.}\cite{Wang2014} & &	
		$^{f}$EAM results of Zhou \emph{et al.}\cite{Zhou2004} \\
		$^{c}$GGA-DFT results of Becquart \emph{et al.}\cite{Becquart2007} & &
		$^{g}$GGA-DFT results of Nguyen-Manh \emph{et al.}\cite{Nguyen-Manh2006}
	\end{tabularx}
\end{table*}
\subsection{Point and planar defects}
Table \ref{tab:purepds} lists the energies of vacancies and self-interstitial atoms (SIAs) in bcc tungsten, essential quantities for the accurate modeling of plasticity. Present MEAM and DFT calculations use a 5$\times$5$\times$5 cubic supercell. Atomic positions are relaxed to 0.01 eV. Geometric details of bcc SIA calculations can be found in Xu and Moriarty\cite{Moriarty1996}. GGA-DFT calculations of Becquart \emph{et al.}\cite{Becquart2007} and the present work indicate the \angbra{111}-dumbell to be the most energetically-favorable self-interstitial, as do the present MEAM potential and F-S potentials of Derlet\cite{Derlet2007} and Ackland\cite{Ackland1987}. Experiments\cite{DiCarlo1969, Okuda1975} and previous MD studies\cite{Zhou2004} found the \angbra{011}-dumbell to be the favored self-interstitial structure in tungsten, but recent work\cite{Amino2016} combining the object kinetic monte carlo (OKMC) method with dislocation loop measurements found OKMC simulations of \angbra{111} interstitials and 1D migration best match experiment. Vacancy formation and migration energies predicted by MEAM compare favorably with present \emph{ab-initio} results and those of Becquart and Domain\cite{Becquart2007} while existing F-S and EAM tungsten potentials are in closer agreement with GGA results of Nguyen-Manh \emph{et al.}\cite{Nguyen-Manh2006}.

\begin{figure}[H]
	\begin{center}
		\includegraphics[width=0.5\textwidth]{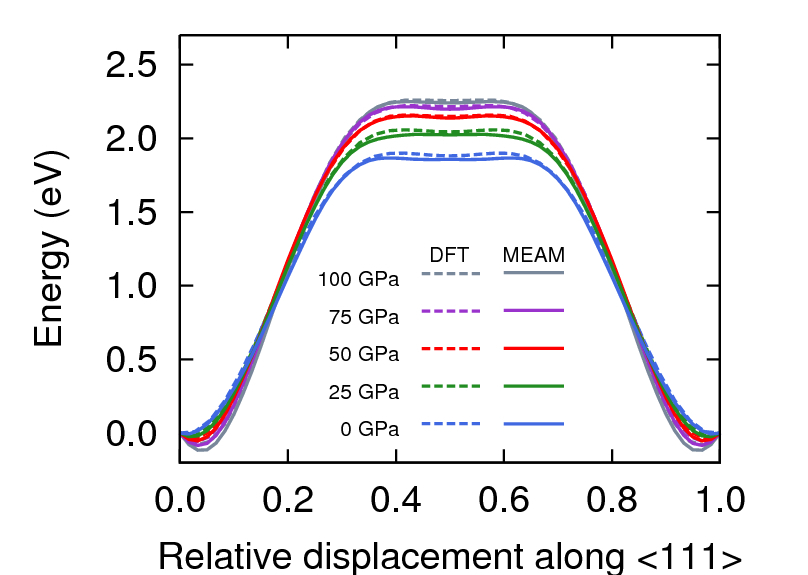}
		\caption{Vacancy migration pathway as calculated in GGA-DFT and MEAM at multiple pressures. The shallow local minimum at \angbra{$\frac{1}{2}\frac{1}{2}\frac{1}{2}$} is predicted by both DFT and MEAM to increase between 0 and 100 GPa, though this effect is non-monotonic in MEAM.} \label{fig:Vac}
	\end{center}	
\end{figure}

\begin{figure}[H]
	\begin{center}
		\includegraphics[width=0.45\textwidth]{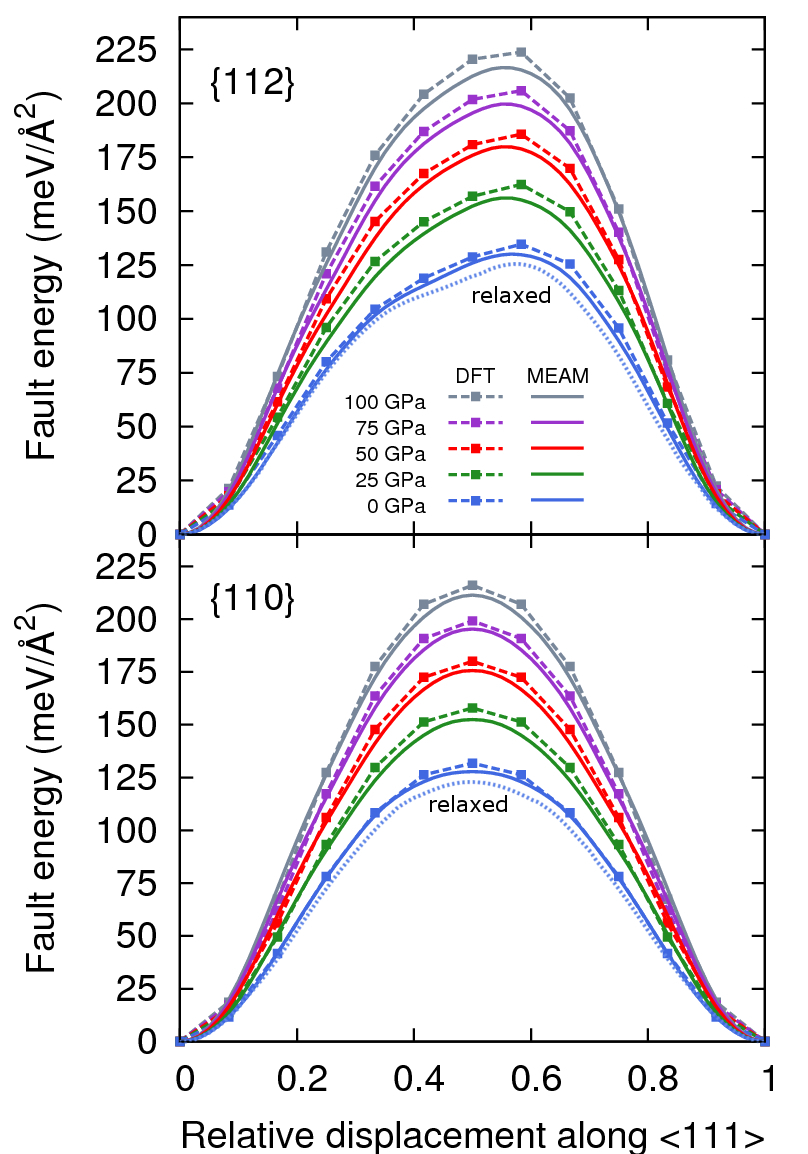} 
		\caption{Unrelaxed low-index generalized stacking fault energies (GSFE) for bcc tungsten. MEAM accurately models the evolution of both \{111\} and \{112\} faults with pressure. Relaxed GSFE curves computed with MEAM are shown as dotted lines for 0 GPa. Relaxation lowers fault energy slightly but does not result in any metastable configurations.} \label{fig:Gamma}
	\end{center}
\end{figure}

\begin{figure}[H]
	\begin{center}
		\includegraphics[width=0.45\textwidth]{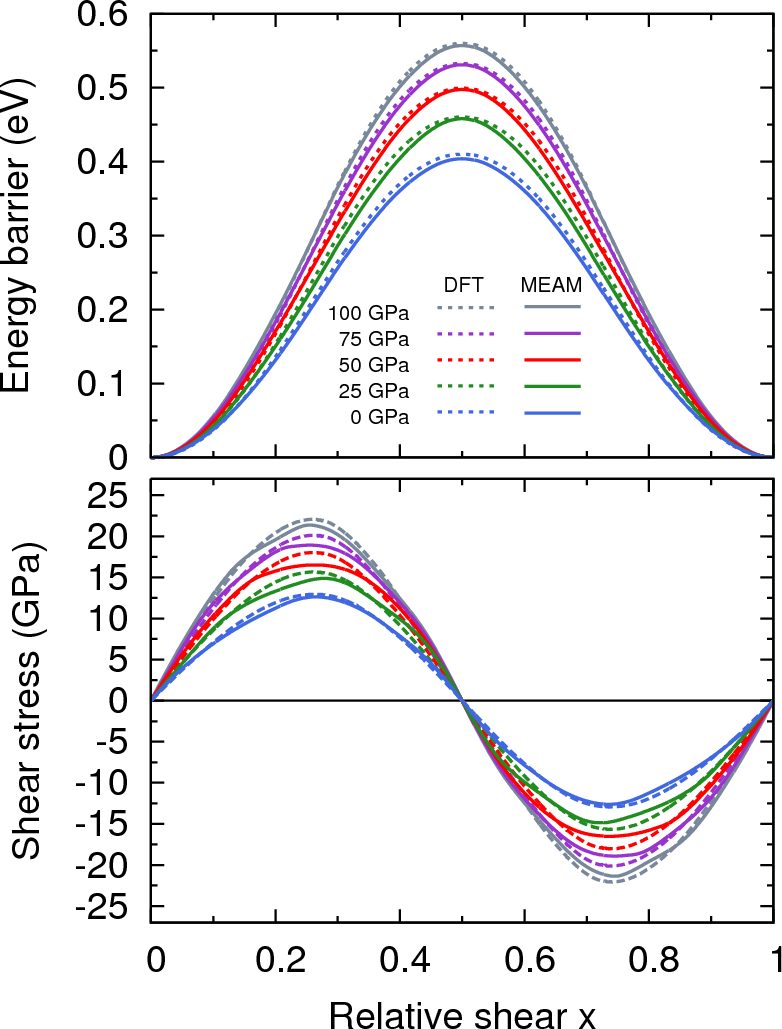} 
		\caption{Ideal shear energy (top) and stress (bottom) for a continuous deformation of a one-atom bcc unit cell corresponding to $(112)[\bar{1}\bar{1}1]$ twinning system as described by Paxton \emph{et al.}\cite{Paxton1991}. MEAM Accurately reproduces the energy barrier and shear stress of this deformation for pressures up to 100 GPa. Small discrepancies in shear stress are found at the inflection points of the energy barrier, which correspond to the two extrema of shear stress at $x = $ 0.25 and $x = $ 0.75.} \label{fig:Ideal}
	\end{center}
\end{figure}

Figure \ref{fig:Vac} presents unrelaxed vacancy migration pathways at five equally-spaced pressures between 0 and 100 GPa. Calculations are performed using a 127-atom 4$\times$4$\times$4 cubic bcc supercell with migration in the \angbra{111} direction. Overall MEAM tracks well with the DFT results; minor discrepancies are found when the vacancy lies near the lattice site and halfway between two lattice sites.
Figure \ref{fig:Gamma} shows unrelaxed generalized stacking fault energies (GSFEs) at five pressures on the $\{112\}$ and $\{110\}$ planes as a function of relative displacement along \angbra{111} for MEAM and DFT. While bcc metals are less prone to stacking-fault formation than their close-packed counterparts, they have been observed in Fe, Nb, W and Mo-35\%Re to exist on $\{112\}$ and $\{110\}$ planes, formed by the dissociation of $\frac{1}{2}$\angbra{111} dislocations\cite{Hirschhorn1963}. Relaxed GSFE curves, computed with MEAM at zero pressure, do not predict the presence of any metastable stacking fault configurations.  At all pressures, MEAM agrees with DFT to within a few meV/\AA$^2$, and thus should be suitable for studying the effect pressure on $\{112\}$\angbra{111} and $\{110\}$\angbra{111} slip systems.
Table \ref{tab:surfs} shows energies and interplanar relaxations of low-index free surfaces. Present calculations employ 48-atom supercells, replicated along the surface normal with an equally-sized vacuum region and periodic boundary conditions. All results presented here predict the \angbra{011} surface to have the lowest energy, followed by \angbra{111}. Finnis-Sinclair potentials\cite{Ackland1987,Wang2014} tend to underestimate the surface energy with respect to GGA-DFT.  The present MEAM potential compares favorably with present \emph{ab-initio} results and those of Vitos \emph{et al.}\cite{Vitos1998} but underestimates the inter-planar relaxation of the high-energy \angbra{100} surface by 50\%. The origin of the discrepancy between GGA-DFT results of Moitra \emph{et al.} and the others is unclear.
\begin{table*}[t]
	\caption{Energy and structural relaxation of low-index free surfaces in bcc tungsten. Surface energies $E$ are given in meV/\AA$^2$ and the changes $\Delta_{12}$ in inter-planar spacing between the first two planes of the surface, where available, are shown in parenthesis in units of percent.} \label{tab:surfs}
	\begin{tabularx}{\textwidth}{l Y Y Y Y Y Y Y Y}
		\hline
		\thline                                   & MEAM$^{a}$ & GGA$^{a}$  & F-S$^{b}$ & GGA$^{c}$ & F-S$^{e}$ & GGA$^{f}$ & AMEAM$^{f}$ & BOP$^{g}$ \\ \hline
		$E_{001}$$\left(\Delta_{12}^{001}\right)$ & 233(-5.7)  & 245(-11.5) & 186(-0.9) & 289       & 183(-0.7) & 487       & 373         & 237(-2.5) \\
		$E_{011}$$\left(\Delta_{12}^{011}\right)$ & 198(-3.8)  & 200(-3.8)  & 159(-1.1) & 249       & 161(-0.5) & 398       & 353         & 163(-1.0) \\
		$E_{111}$$\left(\Delta_{12}^{111}\right)$ & 204(-18.9) & 216(-21.6) & ---       & 278       & ---       & 449       & 314         & ---       \\
		\thline\hline                             
	\end{tabularx}
	\begin{tabularx}{\textwidth}{l | l l}
		$^{a}$MEAM and GGA-DFT results of this work. & &
		$^{e}$Finnis-Sinclair results of Ackland \emph{et al.}\cite{Ackland1987} \\
		$^{b}$Finnis-Sinclair results of Wang \emph{et al.}\cite{Wang2014} & &
		$^{f}$GGA-DFT and AMEAM results of Moitra \emph{et al.}\cite{Moitra2008} \\
		$^{c}$GGA-DFT results of Vitos \emph{et al.}\cite{Vitos1998} & &
		$^{g}$Bond-order potential results of Mrovec \emph{et al.}\cite{Mrovec2007} \\
		$^{d}$Finnis-Sinclair results of Derlet \emph{et al.}\cite{Derlet2007} & & 
		
	\end{tabularx}
\end{table*}
Figure \ref{fig:Ideal} presents the unrelaxed ideal shear stresses and energy barriers for pressures up to 100 GPa. Ideal shear defines the upper limit of stress required to deform a perfect crystal and is fundamental to our current understanding of the strength of materials. Calculations are performed following the methodology of Paxton \emph{et al.}\cite{Paxton1991}, which uses a bcc primitive cell. MEAM accurately reproduces the GGA-DFT results for all pressures, with small discrepancies in shear stress around the extrema.
\section{Transferability of the fitted potential} \label{sec:tra}
Transferability of the fitted potential is demonstrated by application to screw dislocation core structure, deformation twinning in a bicrystal nanorod, and the high-pressure bcc-to-fcc phase transformation.
\subsection{Dislocation core and deformation twinning}
Core structure of the $\frac{1}{2}$\angbra{111} screw dislocation is determined using a cell with lattice directions $\left[1\bar{2}1\right]$, $\left[\bar{1}01\right]$, $\left[111\right]$ and periodic boundary conditions along the dislocation line. The first two lattice vectors are repeated to form a large cell containing 92,277 atoms which are displaced according to the appropriate elastic strain field. The core structure is then determined by relaxing a central region containing 54,396 atoms while the remaining atoms are fixed, ensuring that the correct boundary conditions are satisfied by the long-range anisotropic solution. This methodology is further explained in this group's previous work on Nb\cite{Fellinger2010} and Mo\cite{Park2012}.

\begin{figure}[b]
	\begin{center}
		\includegraphics[width=0.45\textwidth]{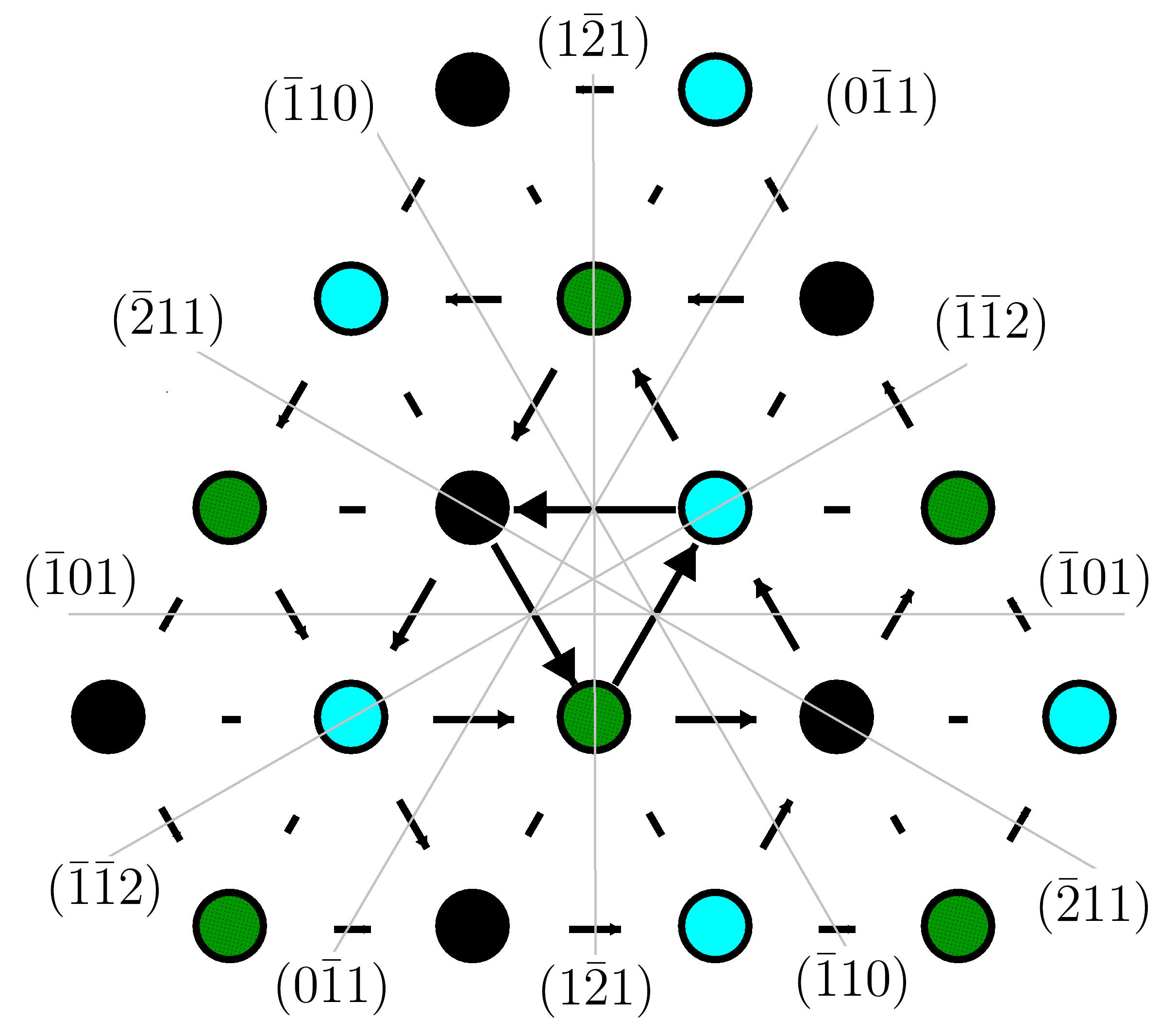} 
		\caption{Differential displacement map of the $\frac{1}{2}$\angbra{111} screw dislocation. MEAM predicts a non-degenerate symmetric core structure consistent with previous bond-order\cite{Mrovec2007} and GGA-DFT\cite{Romaner2010} calculations, whereas existing F-S potentials for tungsten predict a degenerate core\cite{Groger2008}.} \label{fig:discor}
	\end{center}
\end{figure}

Figure \ref{fig:discor} shows a non-degenerate symmetric core structure predicted by MEAM, presented as a differential displacement map\cite{Vitek1970}, is in agreement with results from an existing bond-order potential\cite{Mrovec2007} and DFT-GGA\cite{Romaner2010} calculation for tungsten. Existing F-S potentials predict an asymmetric core\cite{Tian2004,Groger2008}. Our potential is also consistent with the criterion of Duesbery and Vitek\cite{Duesbery1998}, which is based on F-S calculations and states that the $\frac{1}{2}$\angbra{111} screw dislocation in bcc metals will have a symmetric core structure if $\gamma_{\{110\}}(b/3) > 2\gamma_{\{110\}}(b/6)$, where $\gamma_{\{110\}}$ is the relaxed $\{110\}$ $\gamma$-surface and $b = a\sqrt{3}/2$ is the burgers vector magnitude. Relaxed values taken from Figure \ref{fig:Gamma} for MEAM are $\gamma_{\{110\}}(b/3)$ = 100 meV/\AA$^2$ and $\gamma_{\{110\}}(b/6)$ = 39 meV/\AA$^2$.
While dislocation slip is fundamental to plastic deformation of bulk transition metals, twinning has been found to dominate deformation in nanocrystalline Mo, Ta and Fe\cite{Zhu2012}. A recent study\cite{Wang2015} observed deformation twinning and detwinning during uniaxial loading and unloading of a bicrystal nanorod. The Finnis-Sinclair potential of Ackland and Thetford\cite{Ackland1987} was used to model this twinning and detwinning in good agreement with experiment. We simulate this deformation as a challenge for our fitted MEAM potential and to demonstrate transferability to non-equilibrium conditions and consistency with existing models.

\begin{figure*}
	\begin{center}
		\includegraphics[width=1.0\textwidth]{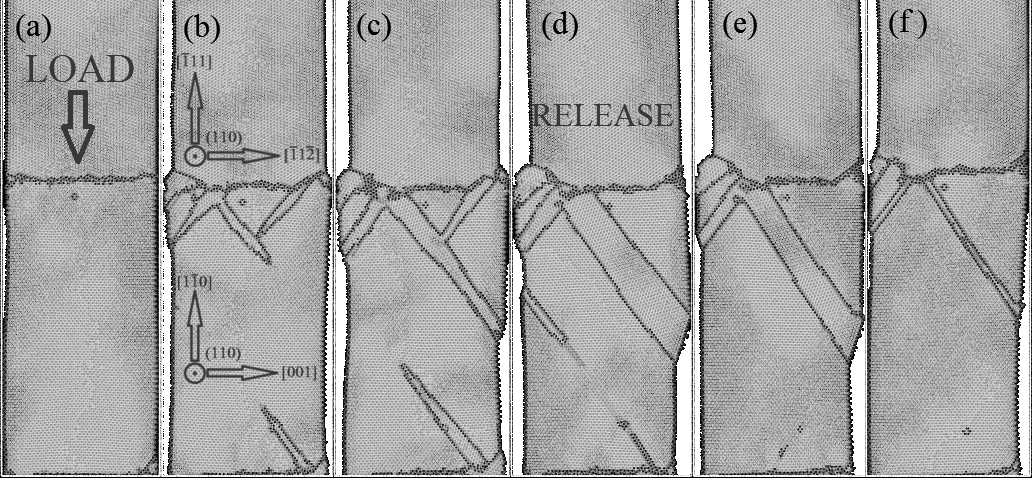}
		\caption{Deformation twinning and detwinning in a tungsten bicrystal nanowire under axial compression at room temperature. Compare to work of Wang \emph{et al.}\cite{Wang2015}. Structure identification was performed using adaptive common neighbor analysis as implemented in {\sc{ovito}}\cite{Stukowski2010,Stukowski2012}. Atoms are color-coded by structure: light (bcc) and dark (none). (a,b,c) Multiple deformation twins of the $\{112\}$\angbra{111} type grow and merge as the rod is compressed by 10\%. (d,e,f) Detwinning occurs as the load is released, recovering the compressive strain. Different shades of gray appear in the bulk because of atomic-level shading in {\sc{ovito}}.} \label{fig:nanorod}
	\end{center}
\end{figure*}

Figure \ref{fig:nanorod} displays cross-sections of a bicrystal tungsten nanorod under uniaxial stress at 300 K. The nanorod is 128 {\AA} in diameter and 510 {\AA} in length, with periodic boundary conditions parallel to the rod axis. A compressive strain of 10 \% is applied from the top of the rod over 20 ps while atomic positions are updated using the Velocity Verlet\cite{Swope1982} integrator and canonical ensemble with 1 fs timestep. The strain is then unloaded over an additional 20 ps. Multiple $\{112\}$\angbra{111} deformation twins can be seen in Figure \ref{fig:nanorod}(a) through (c) to nucleate at the grain boundary and grow with increasing stress. At full loading, strain is accommodated primarily by a single large deformation twin extending from the grain boundary to the rod surface. During unloading the accumulated strain is released by detwinning as can be seen in panels (d) through (f). This deformation behavior is nearly identical to the results of Wang \emph{et al.}\cite{Wang2015}, indicating the transferability of the fitted MEAM potential to modelling tungsten nanostructures and consistency with the successes of previously published potentials\cite{Ackland1987, Wang2015}. Given that the F-S potential of Ackland and Thetford predicts an asymmetric core structure but accurately describes nanorod deformation\cite{Wang2015}, our current MEAM potential is well suited to study the interplay of deformation twinning and dislocation-induced plasticity in tungsten.
\subsection{High-pressure phase transformation}
Finally, we investigate the transformation of tungsten to fcc. Theoretical studies have predicted that bcc tungsten becomes unstable with respect to close-packed fcc and hcp phases at extreme pressures\cite{Einarsdotter1997} and under the conditions of strong electronic excitation during laser irradiation\cite{Giret2014}, for which a $T_{e}$-dependent interatomic potential was developed to study the transition\cite{Murphy2015}. To the authors' knowledge, fcc tungsten has only been observed in thin films formed by sputter deposition between 200 and 400 $^{\circ}C$ on glass, mica and rock-salt substrates\cite{Chopra1967}. The predicted zero-pressure lattice constants of fcc tungsten for MEAM and DFT are 4.049 {\AA} and 4.044 {\AA}, respectively, while Chopra \emph{et al.}\cite{Chopra1967} found an fcc lattice constant of 4.13 {\AA} in the aforementioned tungsten films.  Here we consider the stabilization of fcc over bcc at high pressures, some accessible via diamond-anvil experiments.

\begin{figure*}
	\includegraphics[width=1.0\textwidth]{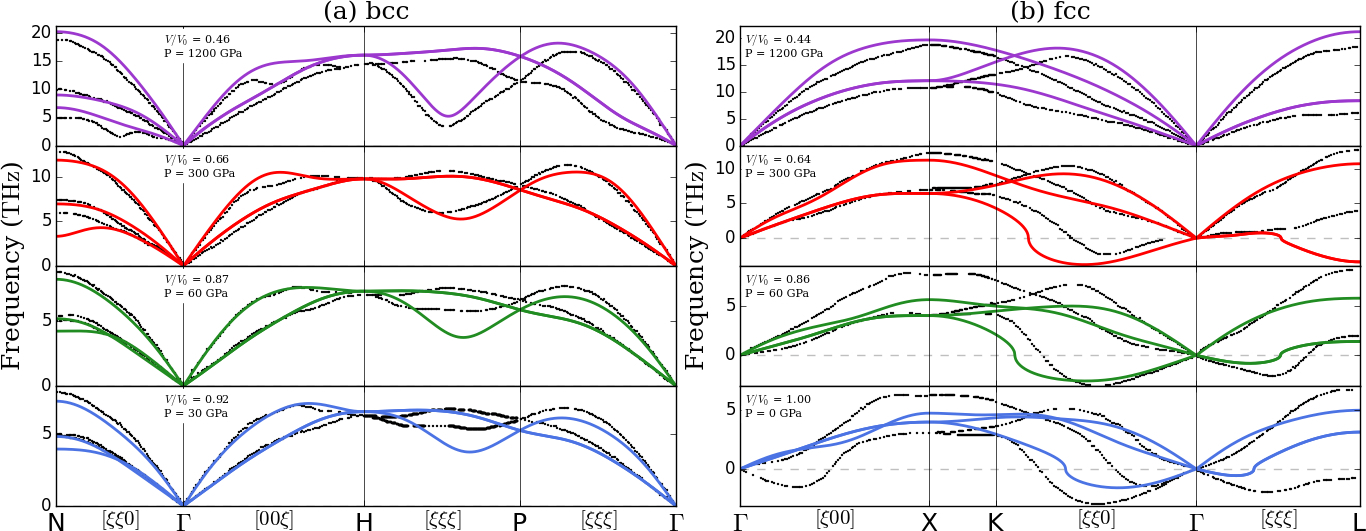}
	\caption{MEAM (solid) Phonon dispersions at various pressures compared with LDA-DFT (dashed) work of Einarsdotter \emph{et al.}\cite{Einarsdotter1997} (a) bcc: MEAM is consistent DFT even at the extreme pressure of 1200 GPa, but underestimates softening rate of the L-$\frac{2}{3}$[111] ($\omega$) phonon and predicts an anomalous softening of the T-$\frac{1}{2}$[110] phonon with increasing pressure. (b) fcc: with the exception of the soft T-[$\xi\xi$0] and T-[$\xi\xi\xi$] modes, MEAM dispersion at zero pressure diverges considerably from that of DFT. MEAM also underestimates the rate of stabilization of the soft modes with respect to DFT work. However at extreme pressures where fcc is thermodynamically stable, MEAM dispersion agrees closely with that of DFT. Displayed pressures are computed with MEAM. These dispersions were calculated using the small displacement method as implemented in the Atomic Simulation Environment\cite{Bahn2002}. As usual for phonons, negative values represent imaginary frequencies.} \label{fig:hpphons}
\end{figure*}

Figure \ref{fig:hpphons}(a) compares MEAM phonon dispersions for bcc W at pressures of 30 to 1200 GPa with LDA-DFT results of Einarsdotter \emph{et al.}\cite{Einarsdotter1997}. MEAM force constants are computed using the small-displacement method, implemented in the Atomic Simulation Environment\cite{Bahn2002}, on a 10$\times$10$\times$10 supercell with $\delta = a(P)/100$ where $a(P)$ is the cubic lattice constant at pressure $P$.  LDA-DFT results employed the density-functional linear response method, norm-conserving pseudopotentials, and 5$s$5$p$5$d$6$s$6$p$ valence.  As seen in Figure \ref{fig:Phonons}, MEAM predicts the L-$\frac{2}{3}[111]$ ($\omega$) phonon to have lower frequency compared with DFT and experiment. This mode softens with increasing pressure, albeit at a lower rate than predicted by LDA calculations. Otherwise MEAM accurately captures the other important features of bcc dispersion up to 1200 GPa. Low-pressure results (30-60 GPa) also compare favorably with the AMEAM results of Zhang and Chen\cite{Zhang2013}.
Figure \ref{fig:hpphons}(b) compares the fcc phonon dispersion predicted by MEAM and LDA-DFT at pressures from 0 to 1200 GPa. At low pressure, where fcc is a highly unfavorable structure, MEAM does not compare well to \emph{ab initio} results but correctly predicts unstable soft modes in the T[$\xi\xi0$] and T[$\xi\xi\xi$] branches. However, the stabilization of these modes with increasing pressure is non-monatonic and particularly anomalous on the T[$\xi\xi\xi$] branch at intermediate pressures. By 1200 GPa, MEAM predicts fcc tungsten to be dynamically stable and shows excellent agreement with the LDA-DFT dispersion.

\begin{figure}
	\begin{center}
		\includegraphics[width=0.47\textwidth]{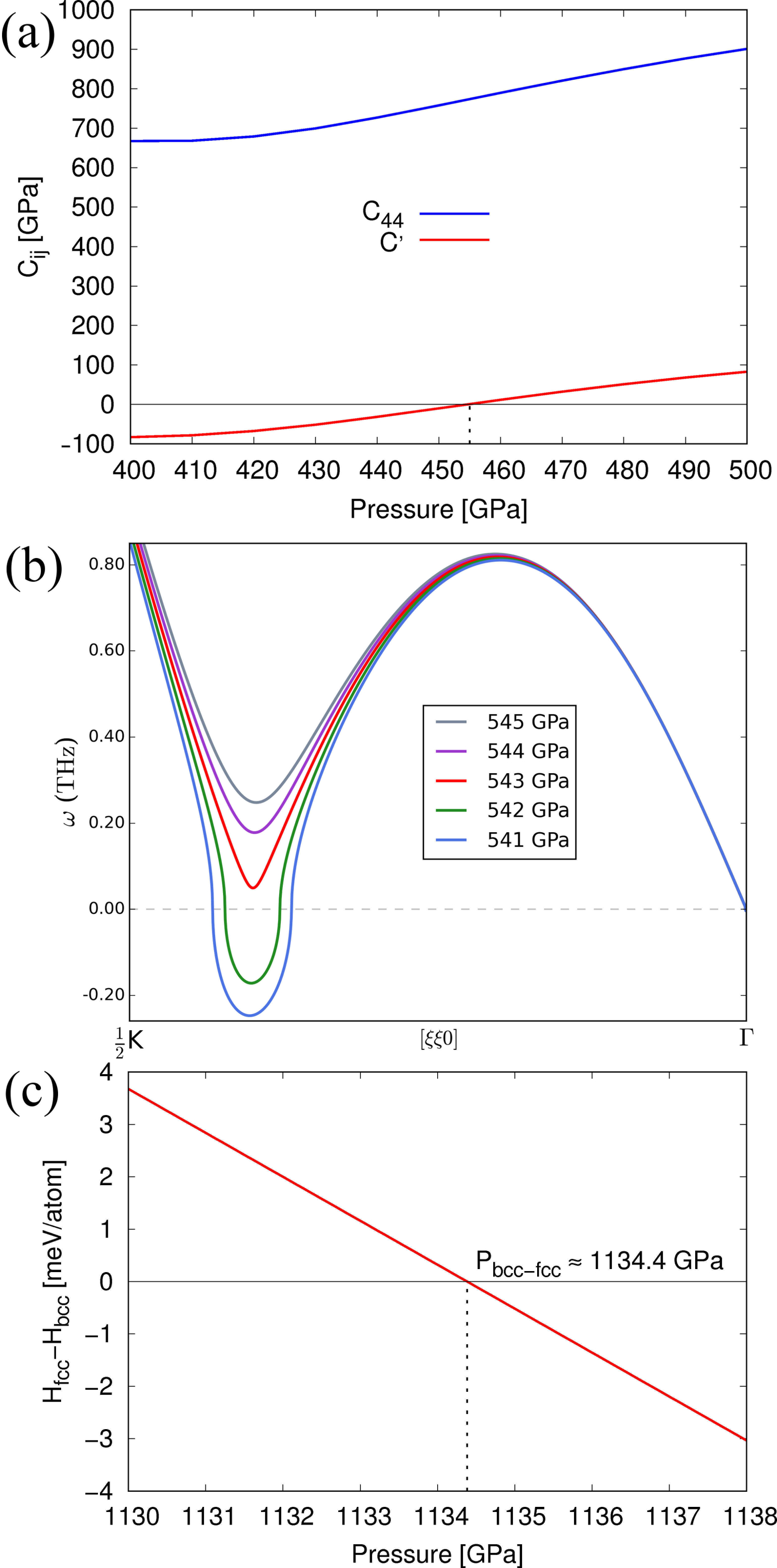}
		\caption{Stability of fcc tungsten. (a) shows the elastic constants C$_{44}$ and C$^{\prime}$ as functions of pressure, demonstrating the elastic stability of fcc tungsten for pressures above 455 GPa. (b) depicts the stabilization of the fcc-T$_{\left[1\bar{1}0\right]}\left[\xi\xi0\right]$ phonon branch with pressure. MEAM predicts that fcc becomes dynamically stable around 543 GPa with the $\xi = 0.4$ mode being the last to stabilize. (c) depicts the enthalpy difference $H_{fcc}(P)-H_{bcc}(P)$ between fcc and bcc as a function of pressure, revealing that despite being dynamically stable, fcc tungsten is not energetically favorable until pressures above 1134 GPa.} \label{fig:HvsP}
	\end{center}
\end{figure}

Figure \ref{fig:HvsP}(a) shows the elasic moduli C$_{44}$ and C$^{\prime}$ between 400 and 500 GPa, where all C$_{ij}$ are positive definite. It can be seen that $\mathrm{C}^{\prime} = \frac{1}{2}\left(\mathrm{C}_{11}-\mathrm{C}_{12}\right)$ is negative for pressures below 455 GPa, reflecting the slope of the T$_{\left[1\bar{1}0\right]}\left[\xi\xi0\right]$ branch arbitrarily close to the $\Gamma$-point. Figure \ref{fig:HvsP}(b) shows this mode for pressures around 540 GPa, where long-wavelength modes are stable but the $\xi$ = 0.40 mode remains unstable. According to MEAM, this mode is the last unstable phonon in any of the considered high-symmetry lines in the Brillouin zone and stabilizes at 543 GPa. However Figure \ref{fig:HvsP}(c), which displays the enthalpy difference $\Delta H = H_{fcc}-H_{bcc}$ versus pressure, shows that the bcc phase remains energetically favorable until about 1134 GPa.
To summarize, present MEAM results are consistent with LDA-DFT predictions of Einarsdotter {\emph{et al.}}\cite{Einarsdotter1997} in that for fcc $C_{44}$ is stable at relatively low pressures, $C^{\prime}$ stabilizes before fcc is thermodynamically favorable, the last phonon mode to become real is the $T_{[1\bar{1}0]}$[$\xi\xi0$] mode at $\xi \approx 0.4$, and that bcc remains energetically favorable until about 1200 GPa. This indicates that the fitted potential is suitable for further study of this high-pressure transformation.
\section{Conclusion} \label{sec:con}
We have developed and applied a novel semi-empirical interatomic potential for tungsten, based on the MEAM and SW formalisms, parameterized using bias-free quintic splines and force-matched to a large database of highly-converged DFT data using an evolutionary global optimization scheme. We have demonstrated accuracy of the fit by reproducing phonon frequencies, compression and thermal-expansion curves, formation energies of unfavorable crystal structures, self-interstitial defects, free surfaces, vacancies, stacking faults and ideal shear at multiple pressures. Transferability of the fitted potential has been demonstrated by  description of the high-pressure bcc to fcc phase transformation, dislocation core structure and deformation twinning and detwinning of a tungsten nanorod. Given the accurate description of both deformation twinning and dislocation structure this potential is more suitable than previous models for studying their interplay. Accuracy of elastic and vibrational properties at high pressures will enable quality shock simulations, and the combination of accurate free-surfaces and non-equilibrium crystal structures should produce reliable descriptions of tungsten nanostructures.
\section*{ACKNOWLEDGMENTS}
Many of the $ab$-$initio$ calculations in our fitting database were performed by Dr. Michael Fellinger. This work was supported by the US Department of Energy under Contract No. DE-FG02-99ER45795 and used computational resources provided by the Ohio Supercomputer Center.
\clearpage

\clearpage
\bibliographystyle{apsrev4-1}
\bibliography{jmeam.bib}{}
\end{document}


	\title{Supplementary Materials: \emph{Ab initio} based empirical potential applied to tungsten at high pressure}
	\maketitle

\section{Parameters of final potential}

\begin{figure}[H]
		\centering
		\includegraphics[width=\textwidth]{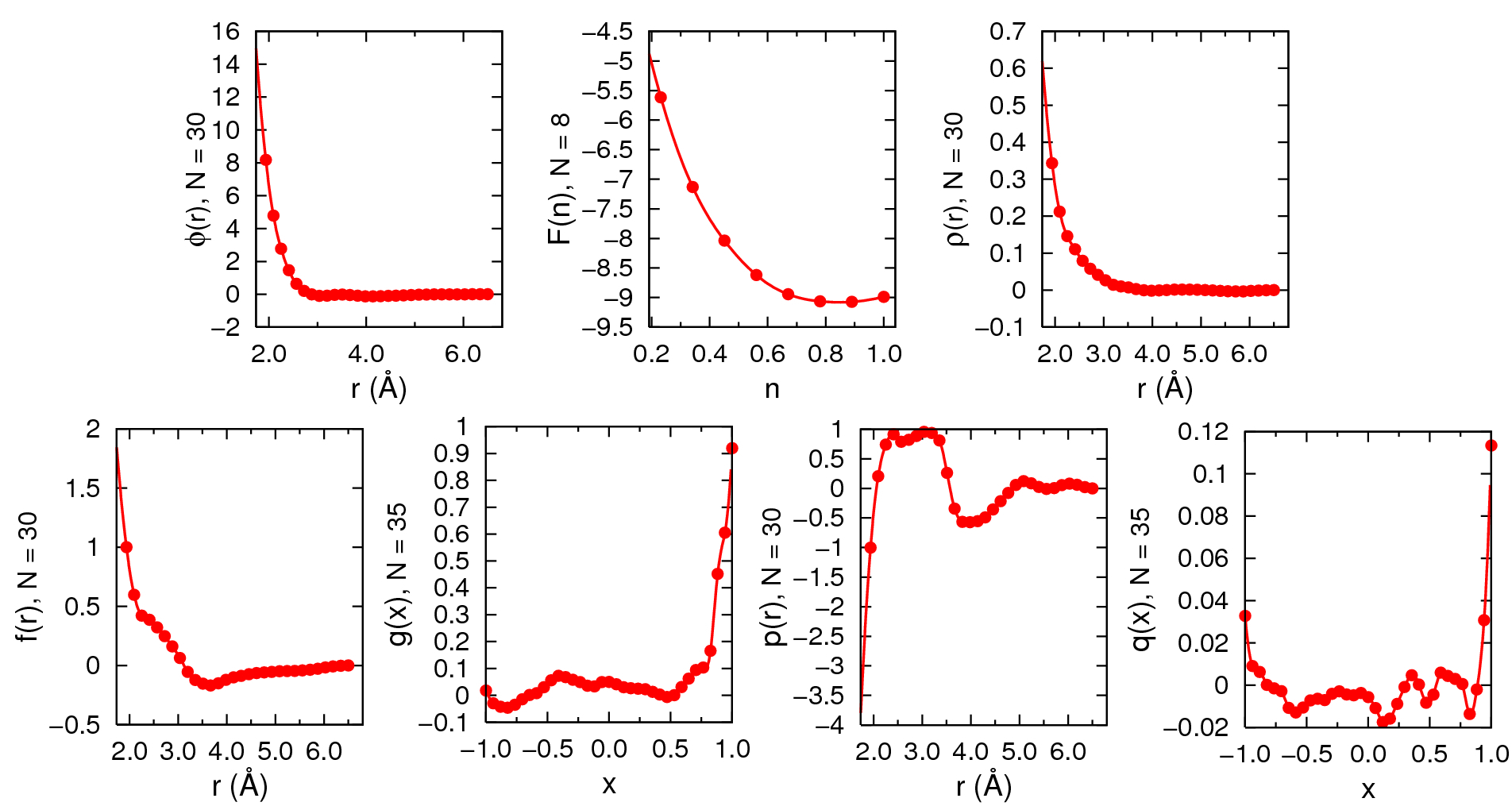}
		\caption{The seven quintic splines comprising the functions of our MEAM model.} \label{fig:splines}
\end{figure}

Figure \ref{fig:splines} plots the seven functions comprising our empirically extended MEAM model. Points represent spline knots whose $y$-values are optimized and lines are the quintic spline interpolants between knots. Values of spline knot $y$-coordinates, ranges of parameters and derivatives at the endpoints are shown in Table \ref{tab:pot}. Given that for all radial functions ($\phi$, $\rho$, $f$ and $p$) the outer spline knot is fixed at zero, the total number of degrees of freedom during optimization is 194.

Source files and the parameter file for the implementation of this model in {\sc{LAMMPS}} are provided with usage instructions in the ``readme" file. These files should be unzipped in the {\sc{LAMMPS}} root directory.

\begin{figure*}
		\begin{table}[H]
			\begin{center}
				\caption{Parameters specifying the seven cubic splines of the MEAM+SW tungsten potential. The first part of the table lists independent variables for each spline function, the bounds of such, and the number of spline knots $N$. In the middle part, values of the spline functions are tabulated for each knot $i$, where $t_{i} = t_{min} + (i-1)(t_{max}-t_{min})/(N-1)$. Lastly, first derivatives are listed for each spline at its endpoint.} \label{tab:pot}
				\resizebox{0.98\textwidth}{!}{%
					\begin{tabularx}{\textwidth}{@{}lYYYYYYY@{}}
						\thline
						\hline
						& $\phi$ & $\rho$ & $U$ & $f$ & $g$ & $p$ & $q$ \\ \hline
						$t$ & $r$ [{\AA}] & $r$ [{\AA}] & $n$ & $r$ [{\AA}] & $\mathrm{cos}(\theta)$ & $r$ [{\AA}] & $\mathrm{cos}(\theta)$ \\
						$t_{min}$ & 1.936 & 1.936 & 0.231451 & 1.936 & -1.00 & 1.936 & -1.00 \\
						$t_{max}$ & 6.50 & 6.50 & 1.00 & 6.50 & 1.00 & 6.50 & 1.00 \\
						$N$ & 30 & 30 & 8 & 30 & 35 & 30 & 35 \\\hline
					\end{tabularx}}
					\renewcommand{\arraystretch}{1.0}
					\resizebox{0.98\textwidth}{!}{%
						\begin{tabularx}{\textwidth}{@{}lYYYYYYY@{}}
							$i$           & $\phi(r_{i})$ [eV]              & $\rho(r_{i})$                       & $U(n)$ [eV]          & $f(r_{i})$                       & $g(x_{i})$          & $p(r_{i})$                       & $q(x_{i})$ [eV]          \\ \hline
							1             & 8.17754794                      & 0.34354505                          & -5.61454815          & 1.00000451                       & 0.01826953          & -1.00355858                      & 0.03275867               \\
							2             & 4.78088386                      & 0.21248938                          & -7.13213530          & 0.59803251                       & -0.02895637         & 0.20528502                       & 0.00908138               \\
							3             & 2.76877268                      & 0.14610410                          & -8.03756015          & 0.42174644                       & -0.04189862         & 0.74087499                       & 0.00625761               \\
							4             & 1.46070080                      & 0.11067492                          & -8.61953525          & 0.38588088                       & -0.04592673         & 0.91254496                       & 0.00010593               \\
							5             & 0.64017812                      & 0.07925502                          & -8.94487218          & 0.32281049                       & -0.03450663         & 0.78648214                       & -0.00154648              \\
							6             & 0.20082462                      & 0.05773259                          & -9.06465740          & 0.24776621                       & -0.01456063         & 0.82030055                       & -0.00285253              \\
							7             & -0.00765327                     & 0.04145698                          & -9.07252007          & 0.16176067                       & 0.00164664          & 0.88605788                       & -0.01075190              \\
							8             & -0.10152106                     & 0.02676237                          & -8.98725594          & 0.06295668                       & 0.00808308          & 0.95485119                       & -0.01295875              \\
							9             & -0.09072809                     & 0.01461594                          &  --                    & -0.05352359                      & 0.03015642          & 0.93639061                       & -0.01060539              \\
							10            & -0.03360000                     & 0.01025421                          & --                     & -0.12298181                      & 0.05630755          & 0.80917294                       & -0.00725918              \\
							11            & -0.01263542                     & 0.00743289                          & --                     & -0.15336078                      & 0.07257101          & 0.26181009                       & -0.00644182              \\
							12            & -0.04609033                     & 0.00312525                          & --                     & -0.16835342                      & 0.06806116          & -0.34277179                      & -0.00705714              \\
							13            & -0.09178435                     & -0.00007454                         & --                     & -0.15045670                      & 0.05831171          & -0.56718711                      & -0.00411671              \\
							14            & -0.12451900                     & -0.00144470                         & --                     & -0.12148391                      & 0.04964783          & -0.57293020                      & -0.00294010              \\
							15            & -0.13252729                     & -0.00080199                         & --                     & -0.10079003                      & 0.03568623          & -0.55459986                      & -0.00439403              \\
							16            & -0.11926969                     & 0.00044550                          & --                     & -0.08713686                      & 0.03376337          & -0.49005332                      & -0.00484268              \\
							17            & -0.10168872                     & 0.00142523                          &  --                    & -0.07416516                      & 0.04927750          & -0.35497080                      & -0.00374903              \\
							18            & -0.08543444                     & 0.00170742                          & --                     & -0.06395072                      & 0.04988678          & -0.21733041                      & -0.00567228              \\
							19            & -0.06753495                     & 0.00143689                          &  --                    & -0.05841828                      & 0.04155554          & -0.07601496                      & -0.01087031              \\
							20            & -0.04517064                     & 0.00096186                          &  --                    & -0.05333600                      & 0.03071994          & 0.05621734                       & -0.01747516              \\
							21            & -0.02724563                     & 0.00011312                          &  --                    & -0.04830354                      & 0.02661144          & 0.12120550                       & -0.01587746              \\
							22            & -0.01558261                     & -0.00083410                         &  --                    & -0.04680635                      & 0.02515616          & 0.08589555                       & -0.00888091              \\
							23            & -0.00722958                     & -0.00184725                         &  --                    & -0.04563719                      & 0.02305014          & 0.02426416                       & -0.00090735              \\
							24            & -0.00526288                     & -0.00288883                         &  --                    & -0.04204248                      & 0.01333036          & -0.00950960                      & 0.00463563               \\
							25            & -0.00738615                     & -0.00349420                         &  --                    & -0.03559011                      & 0.00344994          & 0.00582449                       & 0.00027324               \\
							26            & -0.00672710                     & -0.00318169                         & --                     & -0.02703034                      & -0.00616400         & 0.05373423                       & -0.00829522              \\
							27            & -0.00345596                     & -0.00211846                         & --                     & -0.01676385                      & 0.00075618          & 0.08057652                       & -0.00451289              \\
							28            & -0.00067225                     & -0.00098092                         & --                     & -0.00801562                      & 0.03134535          & 0.06074919                       & 0.00590844               \\
							29            & 0.00018443                      & -0.00028357                         &  --                    & -0.00238667                      & 0.06236685          & 0.02015950                       & 0.00434745               \\
							30            & 0.00000000                      & -0.00000000                         & --                     & 0.00000000                       & 0.09427033          & 0.00000000                       & 0.00293599               \\
							31            &  --                               &  --                                   & --                     &  --                                & 0.10348664          &  --                                & 0.00052140               \\
							32            &  --                               &  --                                   & --                     & --                                 & 0.16594669          &  --                               & -0.01363307              \\
							33            &  --                               &   --                                  & --                     &  --                                & 0.45196039          &  --                                & -0.00205961              \\
							34            &  --                               &   --                                  & --                     & --                                 & 0.60528155          &  --                                & 0.03072894               \\
							35            &  --                               &   --                                  & --                     & --                                 & 0.91946180          &  --                                & 0.11345596               \\ \hline
							
							$i$           & $\phi^{\prime}(r_{i})$ [eV/\AA] & $\rho^{\prime}(r_{i})$ [\AA$^{-1}$] & $U^{\prime}(n)$ [eV] & $f^{\prime}(r_{i})$ [\AA$^{-1}$] & $g^{\prime}(x_{i})$ & $p^{\prime}(r_{i})$ [\AA$^{-1}$] & $q^{\prime}(x_{i})$ [eV] \\ \hline
							1             & -26.28469957                     & -1.06808883                         & -17.15418928          & -3.28496099                      & -1.20699864          & 10.36595022                      & -0.68561309               \\
							N             & 0.00000000                      & 0.00000000                          & 1.22019954          & 0.00000000                       & 8.53609196          & 0.00000000                       & 1.97868195               \\
							\thline\hline &
						\end{tabularx}}
					\end{center}
				\end{table}
\end{figure*}

\section{Algorithms used in fitting procedure}

In this section we present more details about the fitting algorithm. In all flowcharts that follow, ovals represent endpoints of an algorithm, rectangles are ordinary steps and diamonds are decision statements.

\begin{figure}[t]
	\begin{center}
		\includegraphics[width=1.0\textwidth]{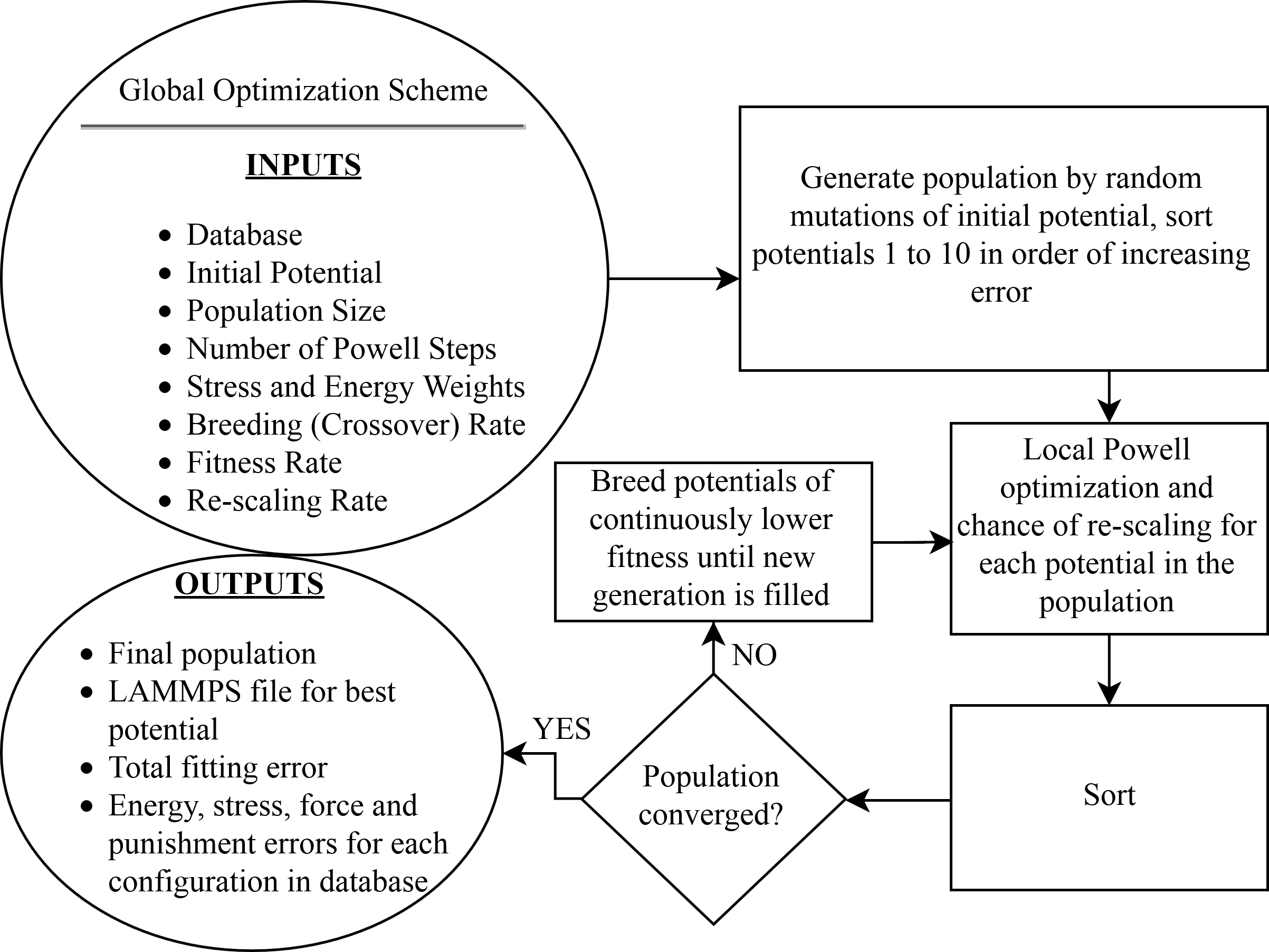}
		\caption{Diagrammatic representation of the global optimization scheme. Ovals represent the beginning and end of the algorithm, with lists of input and output quantities respectively. Rectangles are functional steps in the algorithm and diamonds are "if" statements. The population is deemed "converged" is the change in total error for every member of the population is less than 10$^{-3}$. Partner selection and the breeding algorithm itself are discussed in more detail below.}  \label{fig:meamz}
	\end{center}
\end{figure}

Figure \ref{fig:meamz} presents a flowchart describing the entire algorithm. User-defined inputs include the fitting database, an initial potential (typically with linear functions), breeding population size, number of conjugate direction steps, weighting factors for error calculation, and rates of various stochastic events that occur throughout the algorithm.

\begin{figure}[t]
	\begin{center}
		\includegraphics[width=1.0\textwidth]{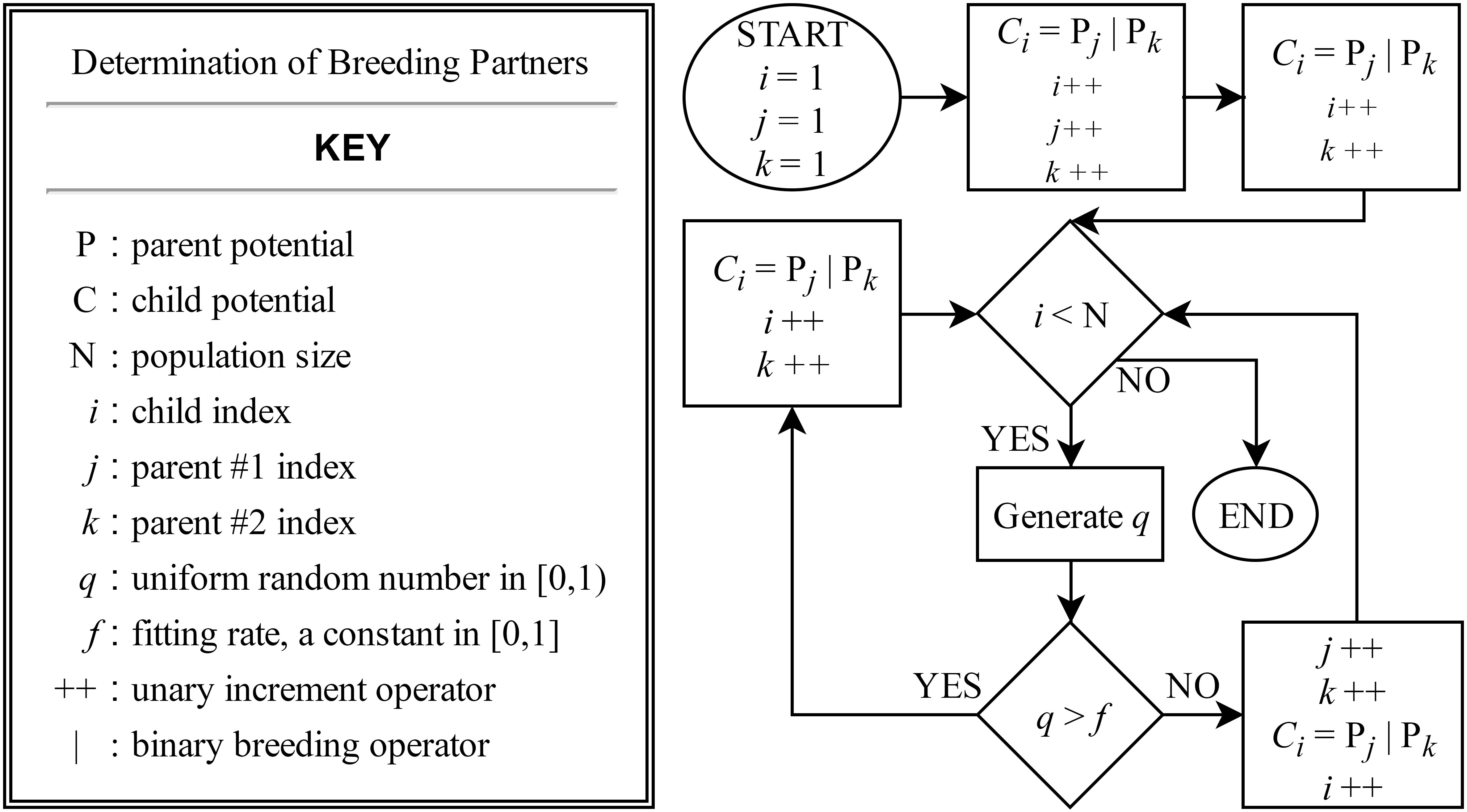}
		\caption{Diagrammatic representation of the algorithm used to determine breeding partners at each step in the evolutionary optimization.}  \label{fig:eugenics}
	\end{center}
\end{figure}

Figure \ref{fig:eugenics} describes the partner-selection algorithm, used to determine breeding partners in a systematic way based on total error. Note that the parents are $pre$-$sorted$ according to total weighted least-squares error.

\begin{figure}[b]
	\begin{center}
		\includegraphics[width=1.0\textwidth]{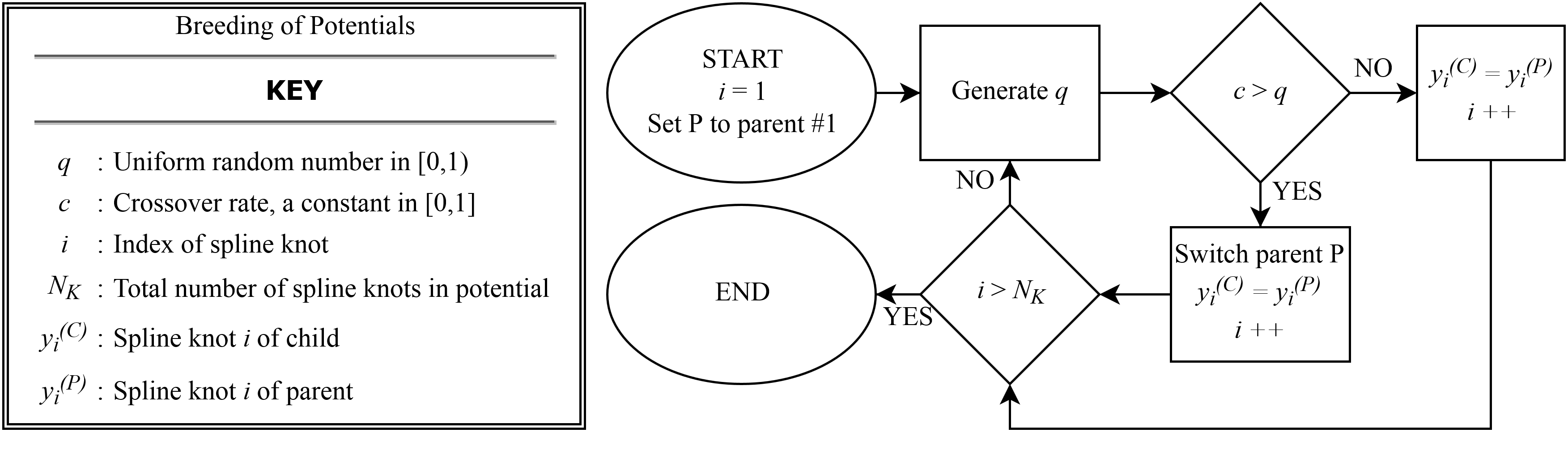}
		\caption{Diagrammatic representation of the algorithm used to breed two parent potentials to form a member of the next generation. Here, parent P represents the ``active" parent from which spline knots are taken.} \label{fig:breeding}
	\end{center}
\end{figure}

Figure \ref{fig:breeding} describes the breeding algorithm. Beginning with the lowest-error parent, there is a probability at each knot that the "genes," or knot values, start being taken from the other parent. This process continues, taking knots from whichever parent $P$ is ``active", until all knots for the child potential have been generated.

